# Possible ferrimagnetism and ferroelectricity of half-substituted rare-earth titanate: a first-principles study on $Y_{0.5}La_{0.5}TiO_3$

Ming An, Huimin zhang, Yakui Weng, Yang Zhang, and Shuai Dong*

Department of Physics, Southeast University, Nanjing 211189, China

**Abstract**

Titanates with the perovskite structure are important functional materials, including ferroelectrics (e.g. $BaTiO_3$) as well as ferromagnetic ones (e.g. $YTiO_3$). Recent theoretical studies predicted even multiferroic states in strained $EuTiO_3$ and titanates superlattices, the former of which has already been experimental confirmed. Here, a first-principles calculation is performed to investigate the structural, magnetic, and electronic properties of yttrium half-substituted $LaTiO_3$. Our results show that the magnetism of $Y_{0.5}La_{0.5}TiO_3$ sensitively depends on the structural details due to the inherent phase competition. The lowest energy state is the ferrimagnetic, giving $0.25\mu_B$/Ti. Furthermore, some configurations of $Y_{0.5}La_{0.5}TiO_3$ show hybrid improper polarizations, which can be significantly affected by magnetism, implying the multiferroic properties. Due to the quenching disorder of substitution, the real $Y_{0.5}La_{0.5}TiO_3$ with random *A*-site ions may present interesting relaxor behaviors.

**Keywords:** titanate, ferrimagnetic, ferroelectricity

**PACS numbers:** PACS 75.85.+t Magnetoelectric effects, multiferroics; 77.84.Cg; PZT ceramics and other titanates; 75.50.Gg Ferrimagnetics

*Corresponding author: sdong@seu.edu.cn

**Introduction**

Strongly correlated electron materials, in the form of transition metal oxides, exhibit many fascinating physical phenomena, e.g. high-$T_C$ superconductivity, colossal magnetoresistivity, and multiferroicity, which have been attracting numerous research attentions [1]. Among various oxides, titanium oxides (titanates) with pseudo-cubic perovskite structures (in a uniform chemical formula $A$TiO$_3$ or $R$TiO$_3$), possess some peculiar physical properties. In titanates, there are enough ferroic-phases, covering ferromagnetic (FM), antiferromagnetic (AFM), ferroelectric (FE), quantum paraelectric, and even the hybrid of these orders. More interestingly, a recent theoretical study predicted strained EuTiO$_3$ to be ferromagnetic and ferroelectric [2], a rare but desired multiferroic property, which was experimentally confirmed later [3]. In addition, an orbital-ordering driven insulating phase was predicted in LaTiO$_3$/SrTiO$_3$ superlattice, which shows ferrimagnetism and ferroelectric [4]. These pioneering works provide promising perspective of titanates to pursuit desired functional properties by careful designs.

Physically, the properties of perovskite titanates depend on the valence of $A$-site cations. For the divalent $A^{2+}$, the Ti$^{4+}$ is usually ferroelectric-active, which can give rise to a proper ferroelectric polarization as in BaTiO$_3$. In contrast, for trivalent $R^{3+}$ (mostly rare earth), the Ti$^{3+}$ is magnetic, with 1 $\mu_B$/Ti moment. The magnetic ground state depends on the size of $R^{3+}$, due to the coupling between the structural-orbital-magnetic degrees of freedom [5-8]. In particular, the G-type AFM (G-AFM) order appears for large $R^{3+}$ (e.g. La$^{3+}$) [5,7], while the FM phase is stable when $R^{3+}$ is small (e.g. Y$^{3+}$) [5,6], as shown in Fig. 1b. This magnetic phase diagram as a function of $R^{3+}$ is somewhat similar to that of $R$MnO$_3$ [9], both of which are full of phase competition, although the detail phases are not identical.

In the present work, we will study Y$_{0.5}$La$_{0.5}$TiO$_3$, namely the half-Y-substituted LaTiO$_3$ or vice versa. There are two purposes. First, according to the plenty experience on manganites, the ionic substitution may generate emergent phases, which are different from the phases of parents [9]. Then it is interesting to ask whether there are any novel new phases hidden in the titanate family? Second, following the recent prediction based on LaTiO$_3$/SrTiO$_3$ superlattice [4], we are curious whether it is possible to achieve multiferroic state in more titanates. In particular, the charge-orbital-ordering, a crucial ingredient in the LaTiO$_3$/SrTiO$_3$ superlattice [4], is a sensitive physical property, which needs proper electronic correlation and special lattice distortions. Is there any alternative choice to avoid these sensitive conditions and to pursuit multiferroicity in an easier way? In our study, since La$^{3+}$ and Y$^{3+}$ are isovalent, no charge ordering will be involved in the Y$_{0.5}$La$_{0.5}$TiO$_3$, leaving a robust Mott insulator with single valent Ti$^{3+}$.

**Model and Method**

LaTiO$_3$ bulk has the orthorhombic structure (space group *Pbnm*) with experimental lattice parameters of $a$=5.636 Å, $b$=5.618 Å, and $c$=7.916 Å [10], as shown in Fig.1a. YTiO$_3$ has the identical structure but slightly different lattice constants, which are $a$=5.338 Å, $b$=5.690 Å, and $c$=7.613 Å [11]. Although sharing the same space group, these two titanates have different magnetic ground states due to different magnitude of GdFeO$_3$-type distortions [2]. According to previous literature, LaTiO$_3$ becomes the G-AFM order below 146 K, while the more-distorted YTiO$_3$ exhibits ferromagnetism below 30 K [5].

Our first-principles calculations are performed using the projector augmented wave approach as implemented in the Vienna *ab initio* simulation package (VASP) [12,13]. The electronic correlation is treated in terms of the generalized gradient approximation (GGA) [14] with Hubbard $U$ imposed on Ti $d$ states, which is set as $U_{\text{eff}}=U-J=3.2$ eV [6-8], using the Dudarev implementation [15]. The optimization and electronic self-consistent interactions are performed using the plane-wave cut-off of 500 eV and the $\Gamma$-centered Monkhorst-Pack $k$-point mesh of 7×7×5. Both the lattice constants and inner atomic positions are fully optimized as the Hellman-Feynman forces are converged to within 10 meV/Å. These parameters are standard choices and have been proved to yield good agreement with experimental data [6-8]. The standard Berry phase method is employed to calculate the ferroelectric polarization [16].

**Results and Discussion**

As a benchmark test, the crystal and magnetic structure of the parent compound LaTiO$_3$ is calculated. First, the fully relaxed lattice of the nonmagnetic state gives $a$=5.714 Å, $b$=5.687 Å, and $c$=8.003 Å, which are close to aforementioned experimental values. The little overestimation of structural size is common when using GGA. Then four magnetic states: FM, A-type AFM (A-AFM), C-type AFM (C-AFM), and G-AFM states, are applied. By comparing their energies, it is demonstrated that the G-type AFM is the stablest state, in agreement with the experimental fact. Besides, YTiO$_3$ has been tested in our previous work [6]. These testes provide a good start point to continue following calculations on Y$_{0.5}$La$_{0.5}$TiO$_3$.

In our study, the Y-substitution is done in the minimal unit cell, which contains four chemical units (i.e. four Ti's). Then for the half-substituted case, there are three independent configurations (denoted as A: rock salt, B: layered, and C: columnar [17]), as shown in Table 1. The structures for these three configurations are fully optimized. The relaxed structures (without magnetism) are slightly different from each other, as summarized in Table 1. According to symmetry analysis, the lattice structures of configurations A and B own a polar point group *mm*2, while C owns a centrosymmetric point group 2/*m*.

Subsequently, the spin-polarized calculations are performed. Both the lattice framework and the internal atomic positions are further relaxed with various magnetic orders. The total energies for all independent configurations are obtained and compared, as shown in Fig. 2a. Not only the aforementioned FM and multiple AFM states are considered, two ferrimagnetic orders (insert of Fig. 2a) are also taken into account considering the lowered symmetry induced by substitutions.

As compared in Fig. 2a, among three configurations, the configuration A has the lowest energy despite the details of magnetic orders. More interestingly, the ferrimagnetic state with the ↑↓↑↑ spin configuration has the lowest energy for both cases A and C, while the A-AFM is the stablest one for the case B (the energy of FM is only a little higher). Thus, the magnetic ground state sensitively depends on the structural details, i.e. the Y-La configuration. And the ferrimagnetic state is quite novel since it does not exist in the phase diagram of unsubstituted $R$TiO$_3$. According to the energy difference between the lowest energy magnetic state and the nonmagnetic state, the ferrimagnetic transition temperature of configuration A can be roughly estimated as ~43 K, with the references of pure YTiO$_3$ and LaTiO$_3$.

This sensibility of magnetism can be reasonably understood considering the phase diagram of $R$TiO$_3$. There is a strong tendency of inherent phase competition, which closely related to the orbital ordering associating with the structural distortions [5-8]. In Y$_{0.5}$La$_{0.5}$TiO$_3$, the distortions are modulated in two aspects: 1) the lattice constants *a-b-c* are compromises between those of LaTiO$_3$ and YTiO$_3$, making the system close to the phase boundary; 2) the local TiO$_6$ octahedron is also tuned, generating multiple values for bond length/angle even within the minimal unit, as shown in Fig. 2b-c. The second point, may further frustrate magnetic exchanges.

The competition between FM and AFM Ti-O-Ti exchanges is determined by the bond structure (length and angle). The underlying driving force is the orbital ordering. Previous theoretical studies showed that the ferromagnetism of YTiO$_3$ is accompanied by a particular orbital order, which is stabilized by $t_{2g}$-$e_g$ hybridization induced by the GdFeO$_3$-type distortion [5]. On one hand, considering the GdFeO$_3$-type distortion, the Y substitution should indeed strengthen the FM tendency in competition. The averaged bond lengths and bond angles of Y$_{0.5}$La$_{0.5}$TO$_3$ are between the original values of parents LaTiO$_3$ and YTiO$_3$, as shown in Fig. 2b-c. On the other hand, the substitution makes anisotropic deformation to TiO$_6$ octahedra, making the bond lengths and bond angles bifurcate (also shown in Fig. 2b-c). With different bond lengths/angles, the exchanges, which determined by the orbital ordering, becomes nonuniform in the system, e.g. giving some ferromagnetic exchanges and some antiferromagnetic ones. Thus, the ferrimagnetic state, as a hybrid state between full ferromagnetic and full antiferromagnetic ones, emerges in the substituted system.

In summary of magnetism, Y$_{0.5}$La$_{0.5}$TiO$_3$ is very possible to show a net magnetic moment due to the ferrimagnetism. The calculated moment is 1 $\mu_B$/u.c. (0.25 $\mu_B$/Ti) for the ideal ferrimagnetic state. Moreover, according to above symmetry analysis shown in Table 1, the substitution can break the space inversion symmetry for configurations A and B, at least in the unit cell scale, which can lead to the local electric dipoles. To confirm this point, the standard Berry phase calculations were performed, and the results are summarized in Table 2. The centrosymmetric phase (e.g. the configuration C) is adopted as the reference state in the Berry phase calculation. It is clear that moderate polarizations indeed exist for the cases A and B, while C is non-polar.

The origin of polarization can be understood as the hybrid improper ferroelectricity, which was first proposed for PbTiO$_3$/SrTiO$_3$ superlattices [18] and thus extended to other superlattices and bulks [19-21]. In our study, all three configurations can be considered as special superlattices stacking along different directions [22]. For example, viewed from a proper direction (Fig. 3), the cases A and B equal to the superlattices with layered modulation of *A*-site cations. Previous studies have confirmed that the modulation of non-polar antiferroelectric distortions (AFD) in $(ABO_3)_1/(A'BO_3)_1$ superlattices oriented along the [001] axis can give rise to an in-plane ferroelectric polarization [23]. Our case B just mimics the [001]-orientated superlattice, while the case A can be considered as the pseudocubic [111]-oriented (the [101] in the *Pbnm* notation) superlattice. Both of these two cases are polar. In contrast, the case C is in analog to the pseudocubic [110]-orientated (the [100] in the *Pbnm* notation) superlattice, which is non-polar according to the symmetry. In this sense, the hybrid improper ferroelectricity predicted in our study is derived from additional lattice distortions induced by mismatch of the different *A*-site cations. Such a geometric ferroelectricity usually can persist to a high temperature above room temperature [18-21].

Although here the polarization is mainly from the structural distortions, the magnetism can still affect the polarization significantly. As shown in Table 2, the polarization changes a lot from nonmagnetic state to magnetic ground state. Then a ferromagnetic transition, e.g. under a strong enough magnetic field, can further suppress the polarization. Especially for the configuration A, the suppression can be up to 86% of the ferrimagnetic one, rendering a promising magnetoelectric effect. Therefore, it is possible to control the ferroelectricity by an applied magnetic field. Even though, the directions of polarizations are not changed upon magnetism. Besides, due to the coexistence of spontaneous polarization and magnetization in the current system, the magnetoelectric response should not be simply linear.

Considering the real compounds, the substituted ions will not be so ideally ordered as the three configurations studied in our simulation, which should be distributed partially in random. Thus the inhomogeneity should exist anywhere in the sample due to this quenching disorder, although the configuration A is most energetically preferred. Considering the insulating behavior (no itinerant electron/hole), the physical properties of real material will reflect averaged effects of all these three structures as well as other possible configurations as sketched in Fig. 4. In this sense, the net magnetization remains expectable considering the ferrimagnetism (cases A and C). Meanwhile, due to the random orientation of electric polarization inside nano-clusters, the net component may be faint. However, the local dipole moments, if polished by electric fields, can give rise to a finite polarization. Due to the inhomogeneity, the critical temperatures for magnetic transition and ferroelectric transition may be broad in two ranges, instead of two points, as in relaxor ferroelectrics [24]. In addition, glassy-like magnetic and ferroelectric behaviors may exist, as observed in phase separated manganites.

**Summary**

In summary, the lattice, magnetic, and electronic structure of Y half-substituted $LaTiO_3$ have been studied using the GGA+*U* method. The magnetic ground state sensitively depends on the structural distortion induced by substitution, and shows probable ferrimagnetism with a moderate net spin moment. In addition, the ferroelectricity is predicted from symmetry analysis and confirmed by the calculation. This polarization can be seen as the hybrid improper ferroelectricity, and can be modulated by magnetism, showing a promising magnetoelectric effect. Our work provides a simple route to realize multiferroicity in titanates.

Work was supported in National Natural Science Foundation of China (Grant Nos. 11274060 and 51322206).

**Figures**

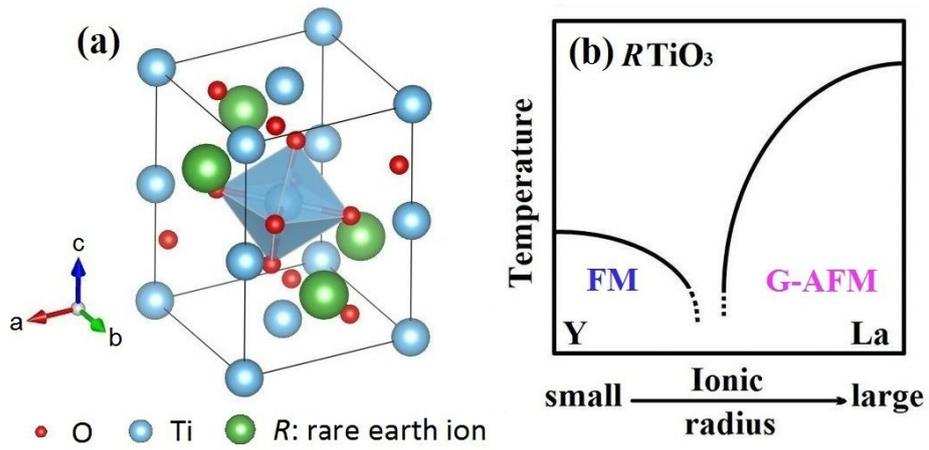

**Figure 1**. (Colour on-line) (a) Schematic structure of $R$TiO$_3$ unit cell. (b) Schematic phase diagram of $R$TiO$_3$ family.

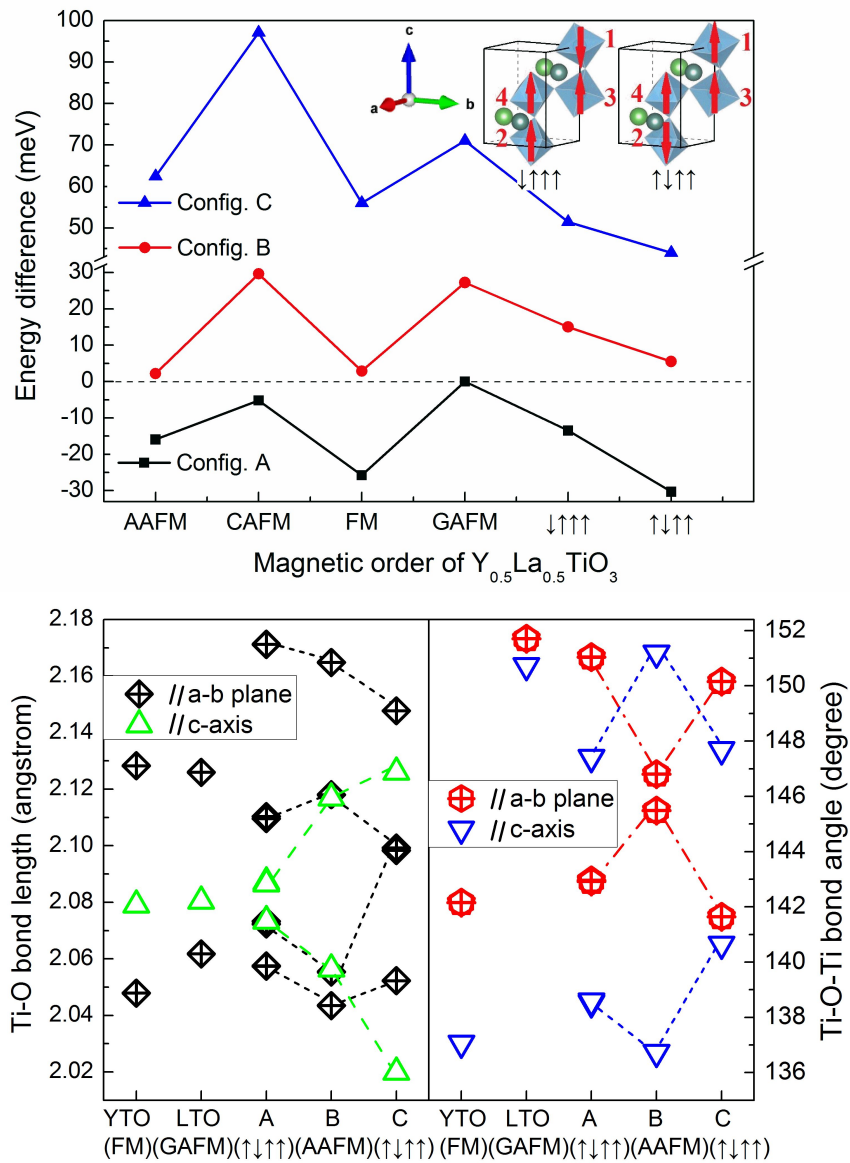

**Figure 2.** (Colour on-line) (a) The total energy of $Y_{0.5}La_{0.5}TiO_3$ in three configurations (A, B, and C) as a function of different magnetic orders including AFM, C-AFM, FM, G-AFM, and two types of ferrimagnetism denoted as ↓↑↑↑ and ↑↓↑↑. The total energies shown here refer to the value of G-AFM state of configuration A for a minimal unit cell (20 atoms) (b) Ti-O bond lengths and (c) Ti-O-Ti bond angles inside and outside *ab*-plane of local $TiO_6$ octahedron in optimized structures with stablest magnetism.

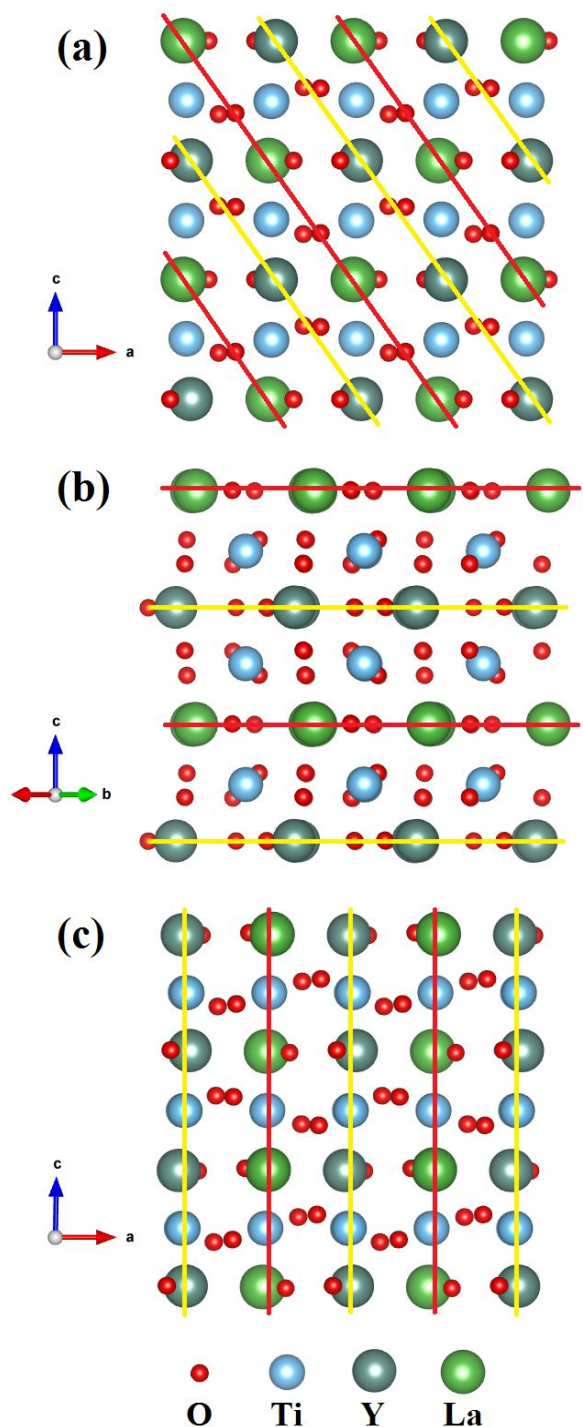

**Figure 3**. (Colour on-line) The superlattice-like structures of $Y_{0.5}La_{0.5}TiO_3$. (a) For the configuration A, it is stacked along the [101] axis in *Pbnm* notation (or the [111] axis in pseudocubic framework). (b) For the configuration B, it is along the [001] direction. (c) For the configuration C, it is along the [100] axis in *Pbnm* notation (or the [110] axis in pseudocubic framework).

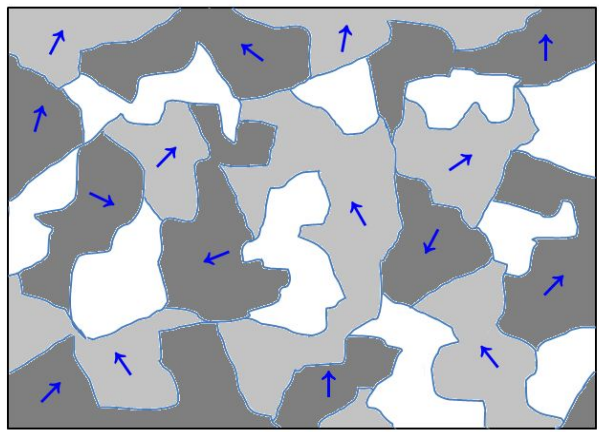

**Figure 4.** Schematic of real compounds with three configurations distributed randomly in nano-scale. Here three types of "domains" (corresponding to the three configurations studied in the present work) are shown although there may exist more possible configurations, some of which possess ferrimagnetism and/or ferroelectricity as indicated by arrows. These local magnetic moments and ferroelectric dipoles can be magnetized and polarized by external magnetic and electric fields, respectively.

**Tables**

**Table 1.** The relaxed lattice structure of $Y_{0.5}La_{0.5}TiO_3$ in three configurations with and without magnetism. Symmetry groups of those structures are also listed here.

| Configuration | A | | B | | C | |
|---|---|---|---|---|---|---|
| 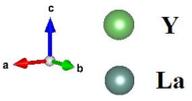 | 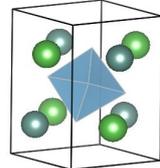 | | 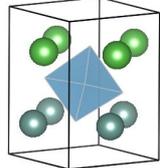 | | 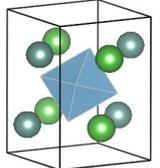 | |
| Relaxed lattice constants (Å) | nonmagnetic | ferrimagnetic | nonmagnetic | A-AFM | nonmagnetic | ferrimagnetic |
| | $a = 5.580$ $b = 5.710$ $c = 7.857$ | $a = 5.552$ $b = 5.841$ $c = 7.850$ | $a = 5.536$ $b = 5.719$ $c = 7.875$ | $a = 5.531$ $b = 5.806$ $c = 7.924$ | $a = 5.567$ $b = 5.714$ $c = 7.867$ | $a = 5.550$ $b = 5.794$ $c = 7.925$ |
| Volume (Å³) | 250.347 | 254.575 | 249.336 | 254.430 | 250.226 | 254.750 |
| Space group | $Pmn2_1$ | | $Pmc2_1$ | | $P2_1/m$ | $P\bar{1}$ |
| Point group | $mm2$ | | $mm2$ | | $2/m$ | $\bar{1}$ |

**Table 2.** The ferroelectric polarization of $Y_{0.5}La_{0.5}TiO_3$ calculated using the Berry phase method. For each configuration, three types of structures are adopted: 1) nonmagnetic optimized structure; 2) the stablest magnetic relaxed structure, and 3) ferromagnetic optimized structure. The difference between these values is due to the magnetic striction effect.

| Configuration | | A | B | C |
|---|---|---|---|---|
| Nonmagnetic | Polarization ($\mu C/cm^2$) | 4.02 | 7.01 | 0 |
| | Direction | [100] | [010] | / |
| Stablest magnetic | Polarization ($\mu C/cm^2$) | 1.70 | 2.64 | 0 |
| | Direction | [100] | [010] | / |
| Ferromagnetic | Polarization ($\mu C/cm^2$) | 0.23 | 2.47 | 0 |
| | Direction | [100] | [010] | / |